\documentclass{article}

\usepackage{microtype}
\usepackage{graphicx}
\usepackage{subfigure}
\usepackage{booktabs} %
\usepackage{hyperref}
\usepackage{pdfpages}

\usepackage[accepted]{mlsys2024}
\usepackage[normalem]{ulem} %

\usepackage[utf8]{inputenc} %
\usepackage[T1]{fontenc}    %
\usepackage{hyperref}       %
\usepackage{url}            %

\usepackage{breakurl}

\usepackage{booktabs}       %
\usepackage{amsfonts}       %
\usepackage{nicefrac}       %
\usepackage{microtype}      %
\usepackage{xcolor}         %
\usepackage{tabularx}       %
\usepackage{amsmath}
\usepackage{multirow}
\usepackage[english]{babel}
\usepackage{blindtext}
\usepackage{tikz}
\usepackage{bm}
\usepackage{filecontents}   %
\usepackage{enumitem}
\usepackage{footmisc}
\usepackage{authblk}
\makeatletter
\renewcommand\AB@affilsepx{, \protect\Affilfont}
\makeatother

\usepackage{tablefootnote}
\usepackage{xcolor} %
\definecolor{LightGray}{gray}{0.9}

\usepackage{CJKutf8}

\newcommand*\circled[1]{\tikz[baseline=(char.base)]{
            \node[shape=circle,fill,inner sep=0.5pt] (char) {\textcolor{white}{#1}};}}
 
\usepackage[labelfont=bf,textfont=bf,skip=8pt]{caption}

\usepackage{listings} %
\newcommand{\myname}[1]{\texttt{CloudEval-YAML}}

\begin{document}

\date{}

\twocolumn[
\mlsystitle{CloudEval-YAML: A Practical Benchmark for Cloud Configuration Generation}

\mlsyssetsymbol{equal}{*}

\centering{
\textbf{
Yifei Xu*$^{1,3}$, Yuning Chen*$^{1,4}$, Xumiao Zhang*$^2$, Xianshang Lin$^1$, Pan Hu\dag$^1$, \\Yunfei Ma$^1$, Songwu~Lu$^3$, Wan Du$^4$, Z. Morley Mao$^2$, Ennan Zhai$^1$, Dennis Cai$^1$
\\
\small
$^1$ Alibaba Cloud $^2$ University of Michigan $^3$ UCLA $^4$ UC Merced 
\\
* Co-first authors. \dag Corresponding author.}
\par
} %

\mlsyskeywords{Large Language Model, Benchmark, Cloud}

\vskip 0.1in

\begin{abstract}
Among the thriving ecosystem of cloud computing and the proliferation of Large Language Model (LLM)-based code generation tools, there is a lack of benchmarking for code generation in cloud-native applications. In response to this need, we present \myname{}, a practical benchmark for cloud configuration generation. \myname{} tackles the diversity challenge by focusing on YAML, the de facto standard of numerous cloud-native tools. We develop the \myname{} benchmark with practicality in mind: the dataset consists of hand-written problems with unit tests targeting practical scenarios. We further enhanced the dataset to meet practical needs by rephrasing questions in a concise, abbreviated, and bilingual manner. The dataset consists of 1011 problems that take more than 1200 human hours to complete. To improve practicality during evaluation, we build a scalable evaluation platform for \myname{} that achieves a 20 times speedup over a single machine. To the best of our knowledge, the \myname{} dataset is the first hand-written dataset targeting cloud-native applications. We present an in-depth evaluation of 12 LLMs, leading to a deeper understanding of the problems and LLMs, as well as effective methods to improve task performance and reduce cost. We release the dataset along with the evaluation framework at \href{https://github.com/alibaba/CloudEval-YAML}{\texttt{https://github.com/alibaba/CloudEval-YAML}}.
\end{abstract}
]

\begin{figure*}[t]
    \centering
    \includegraphics[width=\textwidth]{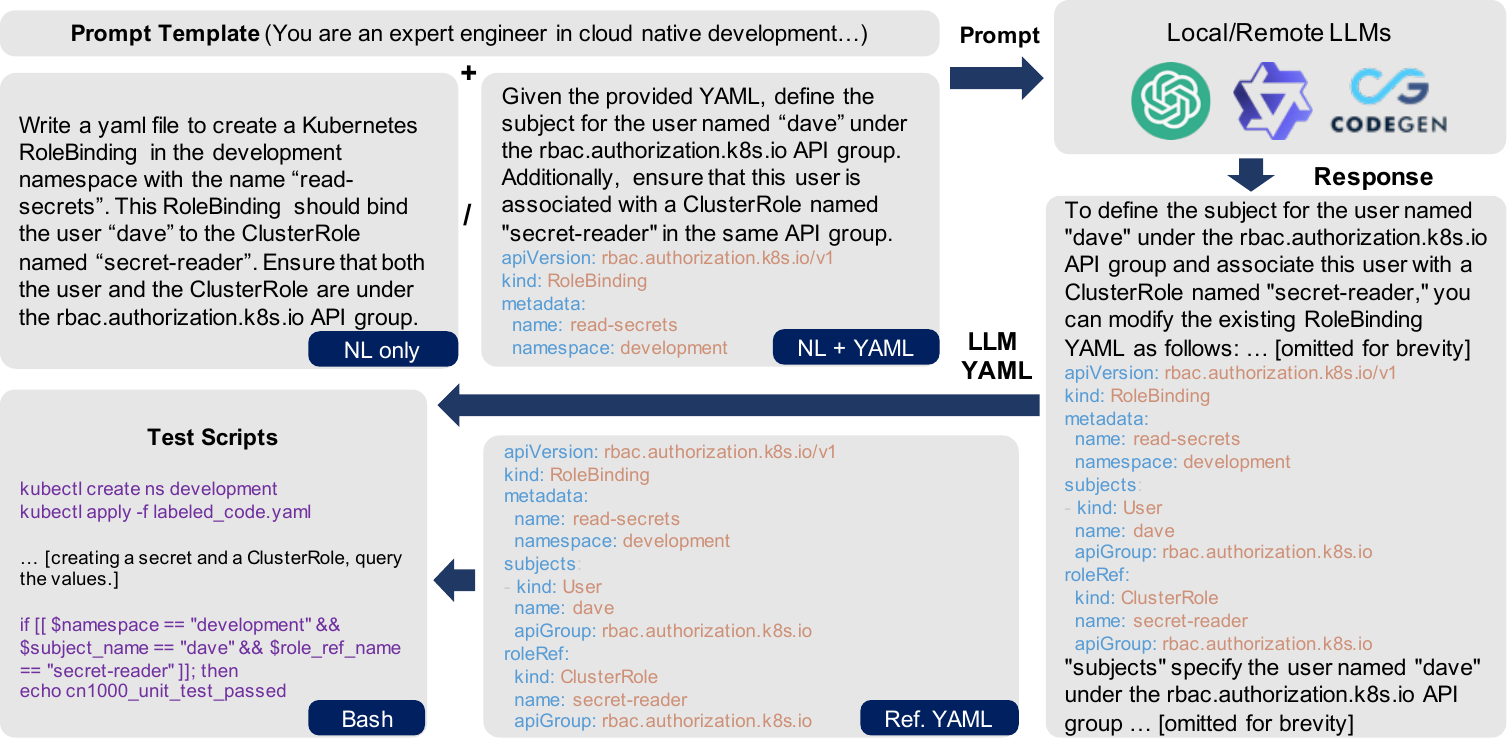}
    \vspace{-14pt}
    \caption{An example problem of the \myname{} dataset, including a problem specification in natural language with an optional sample YAML file as prompt to LLMs, a reference YAML file, and a bash unit test script, to evaluate the YAML output from the LLM.}
    \vspace{-14pt}
    \label{fig:example_problem}
\end{figure*}

\section{Introduction}

Cloud computing is a \$545 billion business~\cite{cloudmarket} with a thriving ecosystem of open-source, cloud-native tools and applications. There are more than 1000 cloud-native applications in the Cloud Native Computing Foundation landscape~\cite{cncf} that have attracted over 4 million Github stars. Many businesses, from small startups to large corporations, rely on cloud services for their operations. On the other hand, recent LLM-based code generation tools such as ChatGPT~\cite{chatgpt, gpt4}, Codex~\cite{codex}, Copilot~\cite{copilot}, and Llama~\cite{touvron2023llama, touvron2023llama2} have been shown to significantly increase the efficiency of software developers~\cite{vaithilingam2022expectation, kabir2023answers}.

Despite the significance of cloud computing and the effectiveness of LLM-based code generation tools, there is a lack of benchmarking for code generation in cloud-native applications. Understanding the performance of various LLM models in the context of cloud applications remains an open question. It is crucial to establish benchmarks that facilitate direct comparisons and foster the evolution of these models and methods.

We attribute the absence of such benchmarks to the following challenges: First, cloud applications are diverse in terms of programming languages and application interfaces, which complicates the construction of a unified benchmark. Second, cloud applications are usually deployed on a large scale, which makes it hard to evaluate the generated code.

In this paper, we present \myname{}, a first-of-its-kind benchmark for cloud configuration generation. It tackles the aforementioned diversity challenge by focusing on YAML, the de facto standard of numerous cloud-native applications. It is both human and machine-readable, thus simplifying the configuration of cloud applications. 90 out of the top 100 most-starred cloud-native applications on GitHub use more than 10 YAML files. We include more detailed statistics in Appendix A.

The \myname{} is designed around practicality, featuring a dataset that consists of 1011 problems and a comprehensive, scalable benchmark framework. The dataset targets realistic problems from a wide range of sources that take over 1200 human hours to complete. Each problem in the dataset includes a question context composed of natural language descriptions and an optional code segment. Additionally, there is a labeled reference YAML file and a unit test script in each problem for evaluation purposes. Figure~\ref{fig:example_problem} shows an illustrative example of a problem.

We further enhance the dataset with practical data augmentation to meet the needs of actual users consisting of simplification and translation rewriting. The simplification involves using concise language with many abbreviations, while translation uses the native language of cloud operation teams. We integrate GPT-4 into the rewriting pipeline to speed up the process, yet we manually review each of them to ensure data quality. 

The \myname{} benchmark also includes a scalable automated evaluation platform with unit tests to ensure functional correctness, as well as a distributed evaluation cluster to score the generated code efficiently. We employ a cloud-based evaluation framework coupled with shared docker image caching. With a cluster of 64 4-core 8GB machines, we managed to complete the evaluation of 1011 problems in 30 minutes, which would take 10 hours on an individual machine. %

We perform an in-depth evaluation on several popular open-source/proprietary language and code generation models with \myname{}, including Llama/Llama2~\cite{touvron2023llama,touvron2023llama2}, Wizardcoder~\cite{luo2023wizardcoder}, Google PaLM-2~\cite{anil2023palm,chowdhery2022palm}, and OpenAI GPT-3.5/GPT-4~\cite{chatgpt,gpt4}. The evaluation includes a comprehensive benchmark of different evaluation metrics, a performance analysis from various perspectives, and experiments with methods to improve generation performance such as multi-sample generation and few-shot prompting. Lastly, we train models that predict unit test results to reduce cost.

The benchmark leads to a series of notable observations, including:
1) Proprietary models outperform open-source models by a large gap.
2) Dedicated code generation models perform poorly on this task.
3) Length of answer and application category are major factors that affect performance.
4) Even the best-performing model makes simple mistakes.
5) Multi-sample generation can provide remarkable gains and can make cheaper models more cost-effective.
6) Few-shot prompts do not show a significant gain on this task.
7) Predicting unit test scores using other metrics can provide an estimation of the ranking.

In summary, our contributions are as follows:
\begin{itemize}[leftmargin=3ex, noitemsep,topsep=0pt]
    \item We present \myname{} benchmark that includes the first hand-written dataset with 1011 practical problems for cloud-scale applications. 
    \item We present the design of a scalable, automated evaluation platform consisting of a computing cluster to evaluate the generated code efficiently for various performance metrics.
    \item We present an in-depth evaluation of 12 LLMs with \myname{}, leading to a deeper understanding of the performance of LLMs in the context of cloud configuration, as well as effective methods to improve task performance and reduce cost.
\end{itemize}

The paper is organized as follows: We first introduce the \myname{} dataset in \S~\ref{sec:dataset}. Then, we detail the \myname{} benchmark platform as well as the evaluation criteria in \S~\ref{sec:platform}, followed by \S~\ref{sec:experiments} where we present the experiments and results. We discuss related works in \S~\ref{sec:related} and conclude the paper in \S~\ref{sec:conclusion}.

\section{The \myname{} Dataset}
\label{sec:dataset}

\subsection{Overall Structure}
The overall structure of \myname{} dataset, illustrated with a simple example, is shown in Figure~\ref{fig:example_problem}. It consists of the following components:
\begin{itemize}[leftmargin=3ex, noitemsep,topsep=0pt]
    \item \textbf{Prompt template}: The prompt template is added to the beginning of each problem to provide context for the model, as well as to specify the output format of the desired answer. We use the same template for all problems, which can be found in Appendix B.
    
    \item \textbf{Natural Language (NL) problem description and optional YAML context}: Within the \textit{myname} dataset, problem descriptions come in two forms: those consisting of natural language only and those paired with a YAML context. The YAML context serves as an input to guide the LLMs in infilling, modifying, or enhancing the functionality of the YAML context.

    \item \textbf{Reference YAML with label}: The dataset also includes a reference YAML file which serves two purposes: First, it can be used as a reference to evaluate the generated YAML file. For example, one can calculate text-level similarities using metrics such as BLEU. However, we find that text-level metrics often fail to consider important properties of YAML such as the fact that object order is not critical. Such oversights can significantly affect the BLEU score. We describe how we tailor the metrics for YAML in \S\ref{sec:platform}. To facilitate YAML-aware comparisons, we include three kinds of labels as comments in the YAML file: 1) wildcard match (labeled with \texttt{\#*}); 2) exact match (by default, no labeling required); and 3) conditional match (e.g., \texttt{\#v in [2,3,4]}). Aside from text-level and YAML-aware metrics, we also use the reference YAML file to facilitate the development and verification of the unit test script.
    
    \item \textbf{Test script}: To benchmark the functional correctness of the generated YAML file, we write automated test scripts to set up the environment and validate the functionality using assertions. For example, if all checkpoints pass, the script outputs \texttt{unit\_test\_passed} to a log file to be further processed and aggregated by the scoring script. The test script also includes a clean-up function ensuring the environment is reset after each test, so the next test can start with a clean environment. 
\end{itemize}

To ensure the authenticity and practicality of the dataset, we collect problems from carefully selected sources including official documentation websites, popular issues from StackOverflow, and highly-ranked blog posts. The problems are hand-picked from these sources based on the following guidelines: 1) The problem should be clearly defined and its purpose should be easily understandable. We avoid problems that are ambiguous in intent or heavily dependent on other YAML files. 2) We collect a diverse set of problems from common cloud applications to ensure our benchmark is comprehensive. We also pick problems of varying difficulty levels, as well as different types of problems that cover common uses of cloud-native tools, as indicated in Table~\ref{tab:statistics}. 3) While it is possible to include scripts like bash, python, or perl in YAML files, we focus on testing the LLM's capability to generate YAML files for cloud applications, rather than creating specific scripts, and avoid problems that require the generation of complex scripts. After hand-picking a problem, we write the problem description and unit tests, then label the YAML file. Part of unit tests are hand-adapted from online sources. More samples of the dataset are shown in Appendix C for an in-depth view of the dataset.

\subsection{Practical Data Augmentation}
\label{subsec:dataset_augment}

\begin{figure*}[t]
    \centering
    \includegraphics[width=\textwidth]{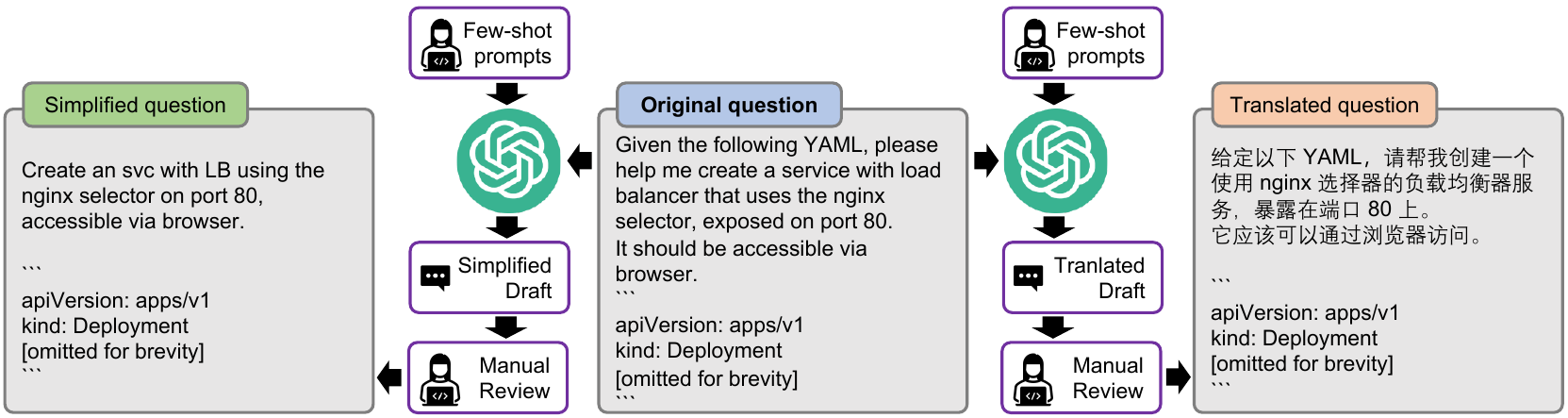}
    \caption{The practical data augmentation framework with examples of a simplified and a translated question.}
    \label{fig:example_augmentation}
\end{figure*}

We conduct a survey with the operation team of a leading cloud provider to align the dataset to representative users. The survey leads to two findings: firstly, actual users often use domain-specific abbreviations and concise language instead of full sentences to express their intentions; secondly, a significant number of users opt to use their native language rather than English. Based on these findings, we apply data augmentation on the original questions to make the dataset more practical, including simplification and translation. We illustrate the process with an example as shown in Figure~\ref{fig:example_augmentation}.

We use GPT-4, a state-of-the-art LLM, to assist the data augmentation process:  \circled{1} we start by writing three examples for simplification and translation. For simplification, we use short and clear language with as many abbreviations as possible. For translation, we use the daily language used by cloud operation teams. The full prompts for both tasks can be found in Appendix D. \circled{2} However, we cannot fully depend on GPT-4 as it can make mistakes. So we manually review the simplified and translated question drafts created by GPT-4 to make sure they keep their original meaning and complexity, ensuring experts can answer them. For the translated drafts, we employ native speakers with cloud development experience to ensure the translated version matches the standard practices of the target developers.

Statistics of the original and augmented datasets are displayed in Table~\ref{tab:augment}. We reduce the average word/token count by 25.7\%/20.9\% respectively via simplification. The translated dataset uses fewer words/tokens, but it's not comparable to the original dataset due to language discrepancy.

\begin{table}[th]
\centering
\caption{\textbf{Statistics of Practical Data Augmentation}}
\label{tab:augment}
\begin{tabular}{cccc}
\toprule
             & Original & Simplified & Translated \\ \hline
Count        & 337      & 337     & 337        \\
Avg. words    & 99.40    & 73.86 \small{(-25.7\%)}   & 57.18      \\ 
Avg. tokens & 508.9     & 402.5 \small{(-20.9\%)}    & 378.5       \\ 
\bottomrule
\end{tabular}
\end{table}

\begin{table*}[t]
\centering
\small
\caption{\textbf{Statistics of \myname{} dataset}}
\label{tab:statistics}
\begin{tabular}{lccccccccc} 
\toprule
\textbf{Statistics} & \multicolumn{6}{c}{\textbf{Kubernetes}} & \textbf{Envoy} & \textbf{Istio} & \textbf{Total / Avg.} \\ 
\cmidrule(l){2-7}
 & pod & daemonset & service & job & deployment & others &  &  & \textbf{/ Max} \\ 
\midrule
Total Problem Count & 48 & 55 & 20 & 19 & 19 & 122 & 41 & 13 & 337 \\
Avg. Question Words & 77.06 & 80.91 & 71.35 & 73.74 & 94.84 & 69.48 & 275.56 & 73.00 & 99.40 \\ 
\midrule
Avg. Lines of Solution & 18.67 & 23.58 & 15.00 & 20.37 & 29.00 & 19.74 & 85.85 & 14.92 & 28.35 \\
Avg. Tokens of Solution & 64.02 & 71.91 & 41.40 & 74.53 & 79.42 & 58.78 & 242.34 & 39.54 & 84.28 \\
Max Tokens of Solution & 150 & 111 & 83 & 163 & 140 & 194 & 531 & 53 & 531 \\ 
\midrule
Avg. Lines of Unit Test & 8.52 & 8.58 & 11.25 & 7.68 & 12.53 & 17.74 & 11.56 & 20.00 & 13.14 \\
\bottomrule
\end{tabular}
\end{table*}

\subsection{Statistics of the \myname{} dataset}

We invest more than 1200 human hours to create \myname{} dataset. The statistics of the dataset are shown in Table~\ref{tab:statistics}. The dataset covers a variety of functionalities in Kubernetes, such as \texttt{pod}, \texttt{daemonset}, \texttt{service}, \texttt{job}, and \texttt{deployment}, while also providing insights into other software like Envoy and Istio, offering a comprehensive overview of each system software's major features and applications. Aside from its wide coverage of real-world applications, the dataset also includes practical problems that are more challenging than other hand-written datasets such as HumanEval~\cite{chen2021evaluating} and MBPP~\cite{austin2021program} in terms of problem/solution length. For example, the average line of solution in this dataset is 28.35, which is 4$\times$ as HumanEval~(6.3) and MBPP~(6.7). We take extra caution in creating the dataset, including limiting the max token count (531) of solutions so it would not exceed the context length of small models, as well as detailed unit tests (13.14 lines on average) to ensure the functional strictness of the test scripts.

\section{The \myname{} Benchmark Platform}
\label{sec:platform}

\begin{figure*}[t]
    \centering
    \includegraphics[width=0.92\textwidth]{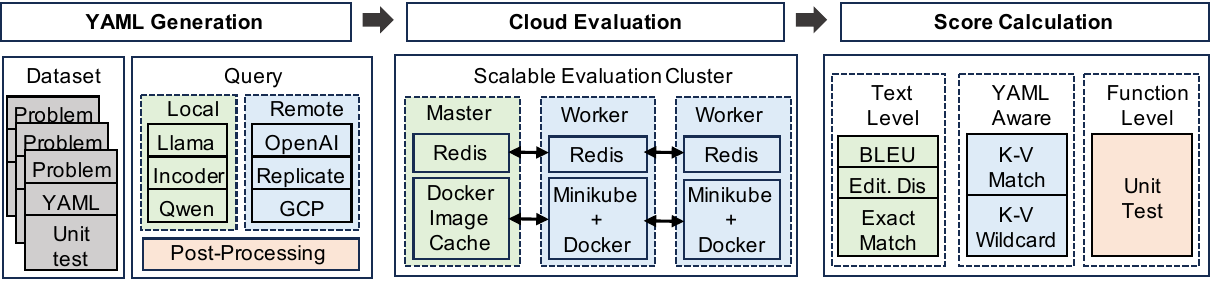}
    \vspace{-0.05in}
    \caption{Workflow of the \myname{} benchmark platform.}
    \vspace{-4pt}
    \label{fig:overall_framework}
\end{figure*}

The overall framework of \myname{} is depicted in Figure \ref{fig:overall_framework}. It is written in 6.3k lines of Python and Bash code, consisting of three major components: YAML generation, cloud evaluation, and score calculation. We first introduce the design of YAML generation and score calculation, and then explain the design of cloud evaluation given the complexity of score calculation. 

\subsection{YAML Answer Generation}

As explained in \S\ref{sec:dataset}, each problem in our dataset includes a file with a problem description. We create prompts for LLMs by combining the prompt template with the problem. These prompts are then processed by the query module to generate a YAML file.

The query module serves as a universal interface for various local and remote models. There are several reasons for doing so. First, it simplifies the differences between different local and remote APIs, offering a consistent interface. This makes it easier to add new models, which only requires adding a query wrapper and a model initializer. Second, it optimizes throughput by using parallel processing to fully utilize the rate-limit and auto-scaling features of Model-as-a-service providers~\footnote{For example, \texttt{replicate.com} provides autoscaling by running multiple copies of a model for no additional cost, as users only pay for the time that \textit{actually running the inference}, rather than idling/booting time. }. As an example, it is straightforward to start 128 raylets with the help of \texttt{ray}~\cite{moritz2018ray} in the query module, which can significantly increase the speed by two orders of magnitude. For local models, the module automatically checks the available GPU memory and adjusts the batch size to speed up inference.

\textbf{Post-processing}: Although we explicitly require LLMs to answer with YAML only, the response often contains text descriptions wrapped around a valid YAML file. We apply the following post-processing policies to extract clean YAML files from such responses:
\begin{itemize}[leftmargin=3ex, noitemsep,topsep=0pt]
    \item Remove all content before the line with the keyword \texttt{Here}, as it is commonly found before the YAML file in responses from several LLMs.
    \item Remove all content before the line with the keyword \texttt{apiVersion:}~(for Kubernetes) or \texttt{static\_resources:}~(for Envoy) since they typically mark the start of a YAML file.
    \item Extract text enclosed by the following delimiters: \texttt{\`{}\`{}\`}, \texttt{<code>} and \texttt{</code>}, \texttt{\textbackslash begin\{code\}} and \texttt{\textbackslash end\{code\}}, \texttt{START SOLUTION} and \texttt{END SOLUTION}.
\end{itemize}

\subsection{Performance Score Calculation}
\label{platform:metrics}

We calculate comprehensive scores using three distinct methods that include a total of 6 metrics to cover different aspects of the generated YAML files. The first method, known as the text-level score, uses metrics such as BLEU, Edit Distance, and Exact Match. The second method, referred to as the YAML-aware score, uses the Key-Value Exact Match and Key-Value Wildcard Match. The third method, the function-level score, uses Unit Tests. Here is how we calculate each:
\begin{itemize}[leftmargin=3ex, noitemsep,topsep=0pt]
\item \textbf{BLEU}: Bilingual Evaluation Understudy~\cite{papineni2002bleu} is a common metric used to evaluate the quality of machine-generated translations. It measures the similarity between the generated YAML and the reference YAML. We use the standard implementation from the NLTK~\cite{bird2009natural}. The BLEU score ranges from 0 to 1, with the higher scores being more desirable.

\item \textbf{Edit Distance}: In some scenarios even imperfect configuration could still be useful if users can fix the error by modifying a few words. The Edit Distance metric is calculated by comparing the number of lines to edit between the generated YAML and the reference YAML using Python standard library \texttt{difflib.Differ}. We scale the edit distance by the size of the reference YAML using \texttt{1 - edit\_distance / len(reference\_YAML)}. As a result, the edit distance score ranges from 0 to 1; the higher, the better.

\item \textbf{Exact Match}: Opposite to edit distance, the exact match score is a very strict metric that requires the generated YAML to be exactly the same as the reference YAML. The output is either 0 (not match) or 1 (exact match).
    
\item \textbf{Key-Value Exact Match}: Different from the exact match that ignores the fact that order doesn't matter in YAML, key-value exact match loads both the generated and reference YAML files into dictionaries and checks if the resulting dictionaries are the same or not, so the output is either 0 (not match) or 1 (exact match).

\item \textbf{Key-Value Wildcard Match}: Similar to the key-value exact match, we also load both YAML files into the dictionary. However, with the help of labeling in the reference YAML file, we can tell what matters and what is not critical. For example, sometimes it is acceptable to start a cluster with \texttt{image:~ubuntu:22.04} or \texttt{ubuntu:20.04}, so the label in reference YAML could be \texttt{image:~ubuntu:22.04 \# v in ['20.04', '22.04']} and either version will be considered correct. We implement this key-value wildcard match using a tree with leaf nodes marked in exact/set/wildcard match and then calculate the IoU (intersection over union) of dictionaries from the generated and reference YAML. The score ranges from 0 to 1; the higher, the better.
    
\item \textbf{Unit Test}: \myname{} ensures the functional correctness of the generated YAML files by running unit test scripts crafted by domain experts. For Kubernetes-focused applications, including Kubernetes and Istio, Minikube~\cite{minikube} offers the capability to set up virtual Kubernetes clusters within a local testing environment. The \texttt{kubectl} command set, the standard tool for managing actual Kubernetes clusters, functions identically on these virtual clusters. It is used for tasks including setting up the environment, applying the YAML files, and monitoring the status.
We use Docker to establish the cluster and perform testing on containers directly for Envoy applications. A unit test is formulated for each issue, tailored to the expected functionality, and it outputs 1 if successful and 0 if not.
\end{itemize}

\subsection{Cloud-based Evaluation Framework}
\label{subsec:cloud_eval}

In contrast to the text-level and YAML-aware scores that take only 21.9 seconds to compute over the entire dataset, running the unit tests in the real environment is very time-consuming: it usually takes several minutes to create the cluster, pull corresponding images, initialize and apply configurations, as well as clear up the environment. This sequential process, especially when scaled to hundreds of problems, becomes a significant bottleneck as it can take more than 10 hours to complete on a single machine. To expedite the evaluation, we employ the following techniques:

\textbf{Scalable Evaluation Cluster:}
We design a scalable evaluation cluster to serve as the unit testing backend. Distinct from the previously mentioned Kubernetes and Docker clusters that run on a local machine, this cluster consists of a master node and multiple worker nodes that span several virtual machines. Central to this system, the master employs a Redis database to manage unit test contexts, inputs, and outputs associated with each problem and benchmark user. Users can dispatch their unit testing jobs to the master, which are claimed by available worker nodes. Subsequently, these workers relay the results back to the master. This paradigm enables automatic parallelization of unit testing, while also ensuring that users can easily monitor progress and access results. Additionally, the distributed design of the evaluation cluster allows for dynamic scaling as needed.

\begin{figure}[t]
    \centering
    \includegraphics[width=0.95\columnwidth]{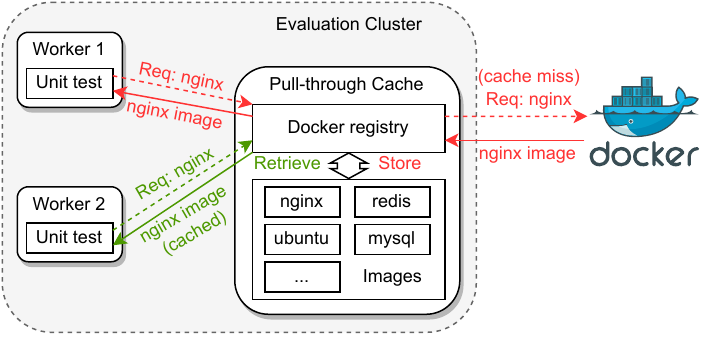}
    \caption{Architecture of shared Docker image caching.}
    \vspace{-4pt}
    \label{fig:docker_proxy}
\end{figure}

\textbf{Shared Docker Image Caching:}
Even with the aforementioned scalable cluster, pulling Docker images from the Dockerhub repeatedly not only consumes excessive internet bandwidth but may also result in an unexpected timeout due to bandwidth variations. To solve this problem, we set up a local docker hub registry that serves as a pull-through cache on the master node, as shown in Figure~\ref{fig:docker_proxy}. This allows workers to share the cached images, removing the need to pull the same image from the Internet if another worker has already downloaded it.

\begin{figure}[t]
    \centering
    \includegraphics[width=0.85\columnwidth]{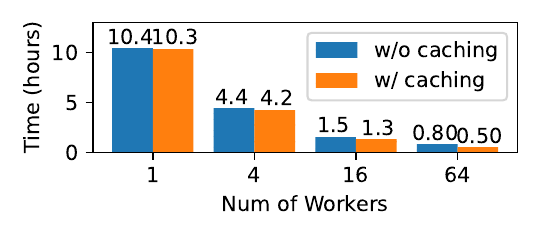}
    \vspace{-1em}
    \caption{Evaluation time over all 1011 problems.}
    \vspace{-4pt}
    \label{fig:exp_cloudeval}
\end{figure}

\textbf{Micro-benchmark of the cloud-based evaluation time}: We run a micro-benchmark on the evaluation time with our cloud-based evaluation framework. The results are shown in Figure~\ref{fig:exp_cloudeval}. We provide the system with 100~Mbps overall bandwidth for internet access. With a cluster consisting of 64 workers, each equipped with 4 CPU cores
and 8GB memory, and shared docker image caching enabled, the evaluation of all 1011 problems is completed in less than 30 minutes. Compared to a single machine, which needs over 10 hours for the same task, this is over $20\times$ speed improvement, with $13\times$ from parallel unit testing and $1.6\times$ from shared Docker image caching.

\subsection{Running Cost of the Benchmark}
\label{subsec:cost}

The main expenses when using the \myname{} benchmark to evaluate are running LLM inference and unit tests in the cloud. Table~\ref{tab:cost} provides a sample cost breakdown of evaluating all 1011 problems for two different inference options (GPT-3.5 API and Llama-7b over \texttt{replicate.com}) and 3 cloud evaluation settings (launching a GCP cluster with 64 standard 4-core 8GB instances, using spot instances to save costs, and using only a single spot instance). The most cost-effective method is using the GPT-3.5 API and a single GCP spot instance, which costs only \$1.31 per run.

\begin{table}[!th]
\centering
\small
\caption{\textbf{Sample Running Cost of the Benchmark in \$.}}
\label{tab:cost}
\newlength{\leftlen}
\settowidth{\leftlen}{Llama-7b: }
\newlength{\middlelen}
\settowidth{\middlelen}{GCP spot $\times$64: }

\begin{tabular}{|l|l|c|} 
\hline
\rule{0pt}{2ex}\textbf{LLM Inference} & \textbf{Cloud Evaluation} & \textbf{Total Cost} \\
\hline
\begin{minipage}[t]{1in}
\vspace{-0.05in}
\begin{itemize}[left=0pt, nosep, label=-]
    \item \makebox[\leftlen][l]{GPT-3.5:} \$0.60
    \item \makebox[\leftlen][l]{Llama-7b:} \$2.90
\end{itemize}
\vspace{0.05in}
\end{minipage} &
\begin{minipage}[t]{1.3in}
\vspace{-0.05in}
\begin{itemize}[left=0pt, nosep, label=-]
    \item \makebox[\middlelen][l]{GCP spot $\times$1:}  \$0.71
    \item \makebox[\middlelen][l]{GCP spot $\times$64:}  \$2.20
    \item \makebox[\middlelen][l]{GCP std. $\times$64:}  \$5.51
\end{itemize}
\vspace{0.05in}
\end{minipage} &
\begin{minipage}[t]{0.5in}
\centering
\vspace{-0.05in}
\$1.31 \\
$\sim$ \\
\hspace*{0.04in}\$8.41
\vspace{0.05in}
\end{minipage}\\
\hline
\end{tabular}
\end{table}

\section{Evaluations on \myname{}}
\label{sec:experiments}

We present an in-depth evaluation of 12 LLMs on the \myname{} in this section, including a comprehensive benchmark of all evaluation metrics and performance analysis from various perspectives including problem/dataset types and failure modes; we also experiment with methods to improve generation performance such as multi-sample generation and few-shot prompting. Lastly, we train models that predict unit test results to reduce cost.

\subsection{Comprehensive Benchmark and Analysis}
\begin{table*}[t]
\centering
\small
\caption{\textbf{Average score of zero-shot benchmark on different models (the higher the better)}}
\label{tab:benchmark}
\begin{tabular}{clccccccccc} 
\toprule
\textbf{Ranking} & \multicolumn{3}{c}{\textbf{Model}} & \multicolumn{3}{c}{\textbf{Text-level Score}} & \multicolumn{2}{c}{\textbf{YAML-Aware Score}} & \textbf{Function-level Score} \\ 
\cmidrule(r){1-1}\cmidrule(r){2-4}\cmidrule(r){5-7}\cmidrule(lr){8-9}\cmidrule(r){10-10}
& Name & Size & \begin{tabular}[c]{@{}c@{}}Open\\ Source\end{tabular} & BLEU & \begin{tabular}[c]{@{}c@{}}Edit\\ Dist.\end{tabular} & \begin{tabular}[c]{@{}c@{}}Exact\\ Match\end{tabular} & \begin{tabular}[c]{@{}c@{}}Key-value\\ Exact\end{tabular} & \begin{tabular}[c]{@{}c@{}}Key-value\\ Wildcard\end{tabular} & Unit Test ~$\downarrow$\\ 
\midrule
1 & GPT-4 & ? & N & \textbf{0.629} & \textbf{0.538} & \textbf{0.092} & \textbf{0.198} & \textbf{0.641} & \textbf{0.515} \\
2 & GPT-3.5 & ? & N & 0.612 & 0.511 & 0.075 & 0.154 & 0.601 & 0.412 \\
3 & PaLM-2-bison~$^1$ & ? & N & 0.537 & 0.432 & 0.040 & 0.092 & 0.506 & 0.322 \\
\midrule
4 & Llama-2-70b-chat & 70B & Y & \textbf{0.355} & \textbf{0.305} & 0.000 & \textbf{0.020} & \textbf{0.276} & \textbf{0.085} \\
5 & Llama-2-13b-chat & 13B & Y & 0.341 & 0.298 & 0.000 & 0.016 & 0.265 & 0.067 \\
6 & Wizardcoder-34b-v1.0 & 34B & Y & 0.238 & 0.247 & \textbf{0.007} & 0.013 & 0.230 & 0.056 \\
7 & Llama-2-7b-chat & 7B & Y & 0.289 & 0.231 & 0.000 & 0.009 & 0.177 & 0.027 \\
8 & Wizardcoder-15b-v1.0 & 15B & Y & 0.217 & 0.255 & 0.002 & 0.002 & 0.226 & 0.026 \\
9 & Llama-7b & 7B & Y & 0.106 & 0.058 & 0.004 & 0.005 & 0.069 & 0.023 \\
10 & Llama-13b-lora & 13B & Y & 0.101 & 0.054 & 0.001 & 0.003 & 0.065 & 0.021 \\
11 & Codellama-7b-instruct & 7B & Y & 0.154 & 0.174 & 0.001 & 0.001 & 0.124 & 0.015 \\
12 & Codellama-13b-instruct & 13B & Y & 0.179 & 0.206 & 0.002 & 0.002 & 0.142 & 0.012 \\
\bottomrule
\end{tabular}

\begin{flushleft}
\hspace{1em} $^1$  The PaLM API supports English only at the time of submission so we averaged the score excluding translated questions.
\end{flushleft}
\end{table*}

\begin{table}[t]
\centering
\small
\caption{\textbf{Number of problems passing unit test on the original and practically augmented datasets}
}
\label{tab:dataset}
\begin{tabular}{lccccccc} 
\toprule
\multicolumn{1}{c}{\textbf{Model}} & \multicolumn{3}{c}{\textbf{Data Set}} \\ 
\cmidrule(r){1-1}\cmidrule(r){2-4}
Name  & Original & \begin{tabular}[c]{@{}c@{}}Simplified\end{tabular} & \begin{tabular}[c]{@{}c@{}}Translated \end{tabular}\\
\midrule
GPT-4  & 179 & 164 \tiny{(-15)} & 178 \tiny{(-1)} \\
GPT-3.5  & 142 & 143 \tiny{(+1)} & 132 \tiny{(-10)} \\
PaLM-2-bison  & 120 & 97 \tiny{(-23)} & N/A $^1$\\ 
\midrule
Llama-2-70b-chat  & 30 & 24 \tiny{(-6)}& 32 \tiny{(+2)} \\
Llama-2-13b-chat  & 26 & 17 \tiny{(-9)}& 25 \tiny{(-1)}\\
Wizardcoder-34b-v1.0  & 24 & 31 \tiny{(+7)}& 2 \tiny{(-22)}\\
Llama-2-7b-chat  & 13 & 9 \tiny{(-4)}& 5 \tiny{(-8)}\\
Wizardcoder-15b-v1.0  & 12 & 11 \tiny{(-1)}& 3 \tiny{(-9)}\\
Llama-7b  & 12 & 7 \tiny{(-5)}& 4 \tiny{(-8)}\\
Llama-13b-lora  & 8 & 9 \tiny{(+1)}& 4 \tiny{(-4)}\\
Codellama-7b-instruct  & 5 & 6 \tiny{(+1)}& 4 \tiny{(-1)}\\
Codellama-13b-instruct  & 5 & 2 \tiny{(-3)}& 5 \tiny{(+0)}\\
\bottomrule
\end{tabular}

\begin{flushleft}
\hspace{1em} $^1$ The PaLM API supports English only.
\end{flushleft}
\end{table}

\begin{figure*}[t]
    \centering
    \includegraphics[width=2\columnwidth]{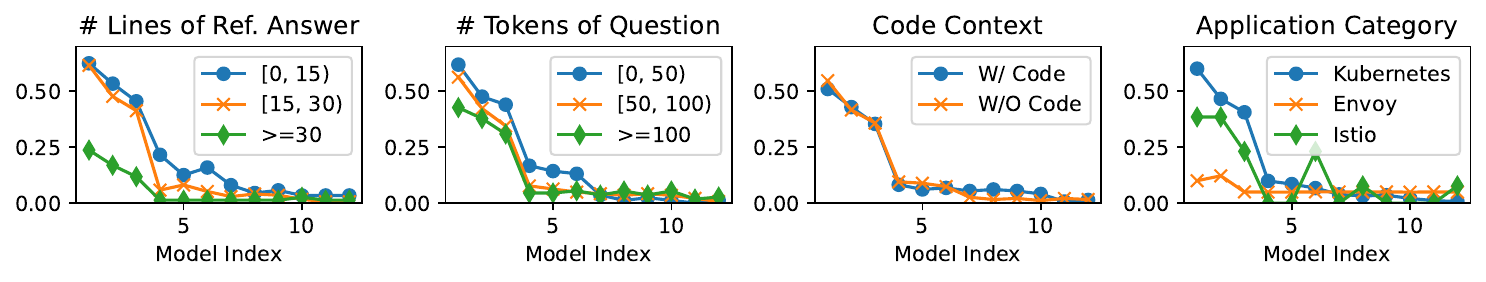}
    \vspace{-4pt}
    \caption{Performance analysis from four perspectives. The x-axis is the model index according to rank in Table~\ref{tab:benchmark}.}
    \vspace{-4pt}
    \label{fig:problem_analysis}
\end{figure*}

We conduct benchmark on both open-source and proprietary models of varying sizes, including GPT-3.5~\cite{chatgpt}, GPT-4~\cite{gpt4}, PaLM 2~\cite{palm2}, Llama~\cite{touvron2023llama}, Llama 2~\cite{touvron2023llama2}, Code Llama~\cite{roziere2023code}, and WizardCoder~\cite{luo2023wizardcoder}. We focus on the "chat" or "instruct" version of models to better fit in the Q/A nature of the configuration generation. The results of our comprehensive benchmark are presented in Table~\ref{tab:benchmark}. \texttt{GPT-4} performs the best in all text-level, YAML-aware and function-level scores. The unit test score is 0.515, meaning that it could pass more than 51\% of the unit tests. Among all the open-source models, \texttt{Llama-2-70b-chat} is the winner that performs best in all metrics other than the \texttt{exact\_match}.

According to the results in Table~\ref{tab:benchmark}, we conclude: \circled{1} Proprietary models such as GPT-3.5 and GPT-4 are way ahead across all metrics, and the gap between them and the best performing open-source models is larger than in similar benchmarks like HumanEval~\cite{chen2021evaluating}. On HumanEval, Llama 2 (70B) is able to achieve a score of 29.9 compared to 48.1 and 67.0 (1.61$\times$ and 2.24$\times$ as Llama 2 (70B)) for GPT-3.5 and GPT-4 respectively. On the unit test score of \myname{}, \texttt{llama-2-70b-chat} scores 0.085 whereas GPT-3.5 and GPT-4 score 0.412 and 0.515, which is 4.84$\times$ and 6.06$\times$ as that of \texttt{llama-2-70b-chat} respectively.  \circled{2} Code LLMs typically perform poorly on \myname{} compared to general LLMs with similar or even smaller sizes in terms of the Unit Test score: \texttt{wizardcoder-34b-v1.0} scores 0.056, while \texttt{llama-2-13b-chat} gets a higher score of 0.067 with less than half the model size. 
It may be related to the dataset used in the fine-tuning process.
However, further investigation is required to uncover the underlying reasons.

\textbf{Performance analysis across various problem types}: The analysis of test results on different types of questions is shown in Figure~\ref{fig:problem_analysis}.
From the analysis, we draw the following conclusions:
\circled{1} While it is not surprising that problems with longer answers are more challenging, we observed a steep decrease in scores for medium-long ([15, 30)) problems. Whether it is due to the emergence of larger models worth future investigation.
\circled{2} The negative correlation between scores and the length of questions is relatively weaker than the length of answers. The longer inputs may indicate more complex requirements, but may also help to provide more context to narrow down the search. 
\circled{3} The presence of a code context doesn't have a substantial influence on the performance most of the time. However, we notice that models with an index from 7 to 10 perform better with code context, indicating that the code may provide some clue about the answer.
\circled{4} Envoy questions are clearly more challenging than Kubernetes and Envoy, which is not surprising due to their longer code lengths compared to other applications as shown in Table~\ref{tab:statistics}.

\textbf{Performance analysis of the practically augmented datasets}: As mentioned in \S{\ref{subsec:dataset_augment}}, the \myname{} dataset comprises both original and simplified/translated questions, tailored to fulfill practical requirements. We compare the performance of these questions, and the result is shown in Table~\ref{tab:dataset}. We focus on unit test scores only for simplicity. The results suggest that: \circled{1} Simplification of problems generally leads to lower performance, but it does not affect large models (such as \texttt{GPT-4} changed from 179 to 164) as much as small models (\texttt{Llama-2-13b-chat}: 26 to 17); \circled{2} Code-specific (\texttt{Wizardcoder-34b-v1.0}: 24 to 2)
and small (\texttt{Llama-2-7b-chat}: 13 to 5) models are severely affected by translation, while large models keep up their performance relatively well(\texttt{Llama-2-70b-chat}: 30 to 32), suggesting the need to fine-tuning small/code-specific models on the bilingual dataset.

\begin{figure}[t]
    \centering
\includegraphics[width=\columnwidth]{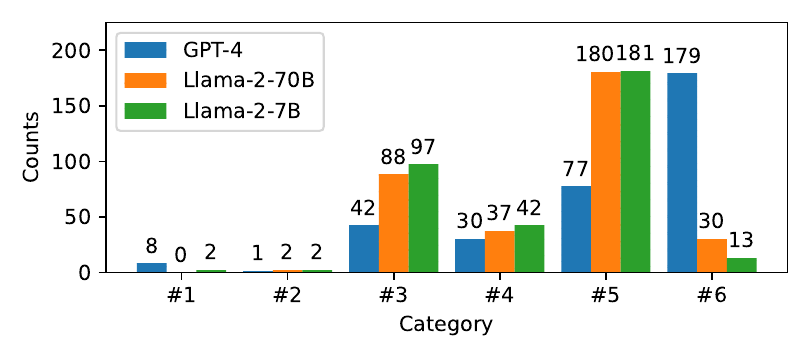}
    \caption{Failure analysis of different models, grouped in 6 modes.}
    \label{fig:failure_mode}
    \vspace{-4pt}
\end{figure}

\textbf{Failure Mode Analysis}: Not all failures are the same: some are close to the correct answer, while others are completely wrong. By analyzing common failure modes, we can understand the weaknesses of each model and find methods to improve the performance. We group the answers into 6 categories, sorted by how close they are to the correct answer: 1) empty or less than 3 lines; 2) longer than 3 lines but does not contain the \texttt{kind} field~\footnote{The field \texttt{kind} exists in most Kubernetes configurations. We search for \texttt{static\_resources} field instead for Envoy configurations.}; 3) contains \texttt{kind} but not a complete YAML file; 4) valid YAML but \texttt{kind} field is incorrect; 5) valid YAML, \texttt{kind} field is correct but the unit test fails; 6) correct YAML that passes the unit test. We analyze the failure modes of the original dataset and the statistics of the result are shown in Figure~\ref{fig:failure_mode}. 

An interesting fact we observed from Figure~\ref{fig:failure_mode} is that \circled{1} the best-performing model GPT-4 makes more Category 1 errors (i.e. simple mistakes) than both Llama2-7B/70B. But given that such errors could be easily filtered, we expect that the performance of GPT-4 could be further improved by implementing a basic format check to filter out such errors and regenerate new ones. On the other hand, \circled{2} both Llama2-7B/70B make a large number of Category 5 errors, suggesting that they are able to get the general idea most of the time, but are not accurate enough to pass unit tests. 

\subsection{Multi-sample Generation}

Sometimes users may not be satisfied with the generated result and might generate additional samples to select the most suitable one. To evaluate the performance of models in such a scenario, we generate multiple samples and evaluate the performance over the original dataset. We select the best-performing open/closed-source models including \texttt{Llama-2-70B, PaLM-2, GPT-3.5} and \texttt{GPT-4} to evaluate their performance with multi-sample generation. We leave parameters that control the randomness of the output to default for proprietary models and set the values of
\texttt{Llama-2-70B} to 0.75/0.9/50 for \texttt{temperature}, \texttt{top\_p} and \texttt{top\_k} respectively. We define pass@k as the number of passed problems where a problem is considered passed if any of its \textit{k} samples passes the unit test~\cite{kulal2019spoc}. 

\begin{figure}[t]
    \centering
    \includegraphics[width=\columnwidth]{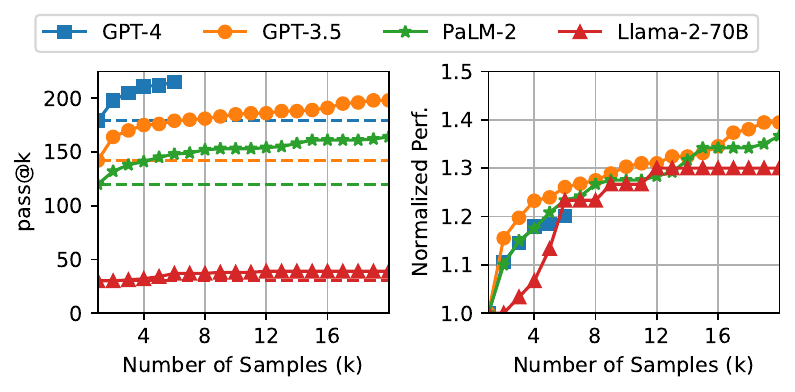}
    \vspace{-1em}
    \caption{Pass@k scores of 4 models in \myname{}.}
    \label{fig:pass@k}
\end{figure}

The result is shown in Figure~\ref{fig:pass@k}\footnote{We run \texttt{GPT-4} for only 6 samples due to the API rate limit.}. The normalized performance indicates an improvement compared with 1-sample results. We observe that: \circled{1} 20-sample generation could improve the unit test score of Llama-2-70B/PaLM-2/GPT-3.5 by 30\%/37\%/39\% respectively, indicating that multi-sample generation could be a good choice to improve the performance \textit{if} there is a unit test or the user can manually select the best result. \circled{2} Although the curves of different models will not cross over each other, it is possible to achieve the same or even better performance with multiple samples. For example, GPT-3.5 with 6 samples could beat GPT-4 with a single sample. Given the 30$\times$ cost differences~\footnote{As of Oct. 1, 2023, the cost for GPT-3.5 turbo with 4k context is \$0.002 per 1k output tokens, while the cost for GPT-4 with 8k context is \$0.06 per 1k tokens.}, it is worth considering using GPT-3.5 with multiple samples as an alternative to GPT-4 with a single sample. Similarly, PalM-2 could reach the GPT-3.5 level after 5 samples, again confirming the effectiveness of this method.

\subsection{Few-shot Prompting}

\begin{table}[t]
\centering
\small
\caption{\textbf{Unit test score on few-shot prompting}}
\vspace{-0.05in}
\label{tab:few-shot}
\begin{tabular}{lcccccccc} 
\toprule
\textbf{Model} & \multicolumn{4}{c}{\textbf{Number of Shots}} \\ 
\cmidrule(r){1-1}\cmidrule(r){2-5}
Name  & 0-shot & \begin{tabular}[c]{@{}c@{}}1-shot\end{tabular} & \begin{tabular}[c]{@{}c@{}}2-shot\end{tabular} & \begin{tabular}[c]{@{}c@{}}3-shot\end{tabular} \\
\midrule
GPT-3.5 & 142 & 150 \tiny{(+8)} & 143 \tiny{(+1)} & 154 \tiny{(+12)} \\
Llama-2-70b-chat  & 30 & 23 \tiny{(-7)} & 26 \tiny{(-4)} & 29 \tiny{(-1)} \\
Llama-2-7b-chat  & 13 & 14 \tiny{(+1)} & 13 \tiny{(+0)} & 15 \tiny{(+2)}  \\
\bottomrule
\end{tabular}
\end{table}

Recent studies have shown the great potential of few-shot prompting across diverse tasks including text classification~\cite{wang2021transprompt, zhang2022prompt, min2021noisy}, logical reasoning~\cite{han2022folio}, code generation~\cite{bareiss2022code, joshi2023repair}. To evaluate the effectiveness of few-shot prompting in cloud configuration generation, we provide the 3 example question-answer pairs as shown in Appendix C, each as one shot, in the prompt and evaluate the unit test scores over the original dataset. The results are shown in Table~\ref{tab:few-shot}.

Surprisingly, our findings suggest that few-shot prompting does not yield significant improvements in unit test scores across varying model sizes. The reason could relate to the heterogeneity of the cloud-native knowledge domain. For example, Kubernetes configurations are distinguished by their \texttt{kind} fields, each having unique formats and essential fields. Given this diversity, few-shot prompting may cover only a limited subset of configuration types, leaving a vast majority unaddressed. Further research is needed to reveal the underlying factors and refine the methodology for cloud-native code generation settings.

\subsection{Predicting Unit Test Results}

\begin{figure}[t]
\centering
\subfigure[Predicted score of all models.]{\includegraphics[width=0.48\columnwidth]{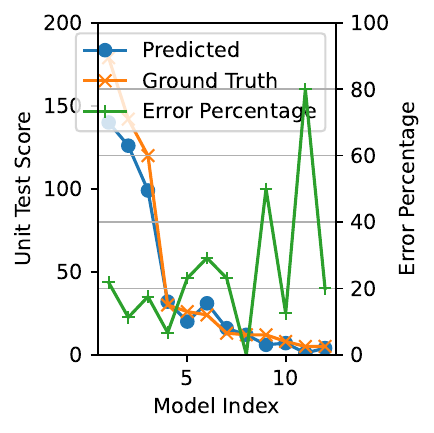}}
\subfigure[SHAP values of different features.]{\includegraphics[width=0.5\columnwidth]{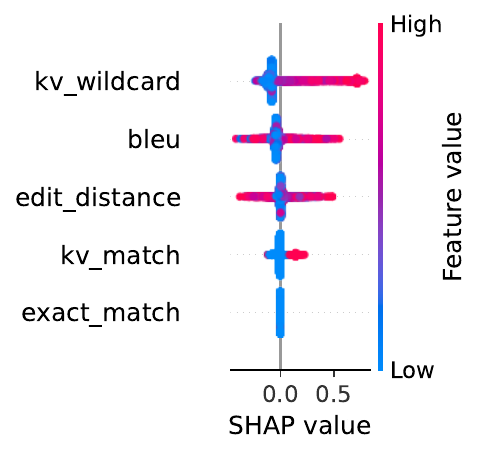}}
\vspace{-0.05in}
\caption{Analysis of unit test classifier using text-level and YAML-aware scores. 
(a) The predicted unit test score of 12 models, sorted by ground-truth unit test values. (b) The SHAP value of each feature in the prediction. }
\vspace{-4pt}
\label{fig:exp_pred_unittest}
\end{figure}

\begin{table*}[t]
\caption{\textbf{Comparison of \myname{} to other benchmarks for code generation.}}
\vspace{-0.05in}
\label{tab:cmp}
\small
\centering
\begin{tabular}{@{}lccccc@{}}
\toprule
\textbf{Dataset} & \textbf{Problem Domain} & \textbf{Special Eval. Metric~$^1$} & \textbf{\# of Problems} & \textbf{Data Source} & \textbf{Natural Lang.}\\ 
\midrule
HumanEval & Python algorithm & {Unit tests} & 164 & \textbf{Hand-written} & EN \\
MBPP & Basic Python & {Unit tests} & 974 & \textbf{Hand-verified} & EN \\
WikiSQL & SQL query & Execution Accuracy & 88k & \textbf{Hand-annotated} & EN \\
\midrule
CodeApex & C++ algorithm & {Unit tests} & 476~$^2$ & Online judge system & \textbf{EN}, \textbf{ZH} \\
MCoNaLa & Python & - & 896 & StackOverflow & \textbf{EN}, \textbf{ES}, \textbf{JA}, \textbf{RU} \\
Lyra & Python w/ embed. SQL & {Code exec./AST} & 2000 & GitHub & \textbf{EN}, \textbf{ZH} \\
\midrule
APPS & Python & {Unit tests} & 10k & Codeforces, Kattis & EN \\
CoNaLa & Python, Java & - & 2879 & StackOverflow & EN \\
Django & Python Django & Human study & 19k & Django codebase & EN \\
Shellcode\_IA32 & Assembly & - & 3200 & shell-storm, Exploit & EN \\
CodeXGLUE & Python, Java & -  & {645k}~$^3$ & Various sources & EN \\
CONCODE & Java classes & - & 100k & GitHub repositories & EN \\
DS-1000 & Python data science & {Unit tests} & 1000 & StackOverflow & EN \\
Ansible & YAML for Ansible & K-V match & 112k & GitHub, GitLab & EN \\
\midrule
\myname{} & YAML for Cloud apps & {Unit tests}, K-V wildcard  & 1011 & \textbf{Hand-written} (337/1011) & \textbf{EN, ZH}\\ 
\bottomrule
\end{tabular}

\begin{flushleft}
\hspace{1em} $^1$ We exclude widely used text-level evaluation metrics such as exact match and BLEU.
\\
\hspace{1em} $^2$ We include the problems of the \texttt{code generation} task only, excluding the \texttt{programming comprehension} task.
\\
\hspace{1em} $^3$ We include the \texttt{text-code} category only, excluding \texttt{code-code}, \texttt{code-text} and \texttt{text-text}.
\end{flushleft}
\vspace{-4pt}
\end{table*}

As we have demonstrated in \S\ref{subsec:cloud_eval} and \S\ref{subsec:cost}, running the unit tests comes with both time and financial costs. Can we spare the costs by predicting unit test scores from text-level and YAML-aware scores?

To answer the question, we collect over 4000 YAML files and their evaluation scores, generated by 12 distinct models. 
Using these scores, we train an XGBoost-based classifier~\cite{chen2016xgboost} to predict whether a specific YAML file would pass the unit test, considering both text-level and YAML-aware scores as input features. We simulate scenarios in which we evaluate new models without running unit tests by masking the target unit scores and training the classifier on the data from the rest 11 models. The result is shown in Figure~\ref{fig:exp_pred_unittest}(a). The relative order could be maintained for most cases by the classifier except for \texttt{wizard-34b} and \texttt{codellama-7b}. However, the relative error could be as high as 80\%, with most errors between 5\% to 30\%. In conclusion, the classifier provides a rough understanding of the ranking, but we still need to run unit tests for more accurate evaluation.

We also perform SHAP (SHapley Additive exPlanations)~\cite{lundberg2020local} to further analyze the significance of each feature. The result is shown in Figure~\ref{fig:exp_pred_unittest}(b), indicating that key-value wildcard match is the most important and accurate feature, while BLEU and Edit distance are important but noisy metrics.

\section{Related works}
\label{sec:related}

We compare \myname{} to other benchmarks as shown in Table~\ref{tab:cmp}.
\myname{} is a hand-written dataset that allows us to customize to specific domains and focus on real-world problems, leading to a more realistic evaluation. There are several other hand-written benchmarks, including HumanEval~\cite{chen2021evaluating}, MBPP~\cite{austin2021program} and
WikiSQL~\cite{zhong2017seq2sql}. HumanEval contains 164 hand-written Python programming problems that benchmark language comprehension, algorithms, and simple mathematics, and MBPP~\cite{austin2021program} contains 974 entry-level Python problems, while WikiSQL~\cite{zhong2017seq2sql} is a hand-annotated benchmark dataset aimed at converting natural language queries into SQL queries. None of these benchmarks is targeting cloud applications.

Aside from that, there are many non-hand written datasets derived from online sources, including APPS~\cite{hendrycks2021measuring}, CoNaLa~\cite{yin2018learning, orlanski2021reading}, Django~\cite{oda2015learning}, Shellcode\_IA32~\cite{liguori2021shellcode_ia32}, CodeXGLUE~\cite{lu2021codexglue}, CONCODE~\cite{iyer2018mapping}, DS-1000~\cite{lai2023ds}, and Ansible-YAML~\cite{pujar2023automated}. Among them, DS-1000~\cite{lai2023ds} is a dataset of Python data science problems collected and adapted from StackOverflow. Ansible-YAML~\cite{pujar2023automated} focuses on the development of Ansible Wisdom, a natural language to Ansible-YAML code generation tool, both works inspire the design of \myname{}, but we choose to focus on the hand-written, realistic problems that focus on real cloud application and are not common in the public datasets.

Some benchmark datasets have been created to support multiple natural languages. CodeApex~\cite{fu2023codeapex} assesses the performance of LLMs involving bilingual prompt strategies. MCoNaLa~\cite{wang2022mconala} is another multilingual dataset modeled off of the methodology from CoNaLa~\cite{yin2018learning} while extending beyond English. Lyra~\cite{liang2021lyra} is a dataset in Python with embedded SQL for code generation tasks. Each program comes with both Chinese and English comments.

\section{Conclusion}
\label{sec:conclusion}
We present the \myname{} benchmark, consisting of the first hand-written dataset that includes more than 1000 realistic problems for YAML configuration generation in cloud applications, accompanied by a complete end-to-end evaluation platform and functional correctness evaluation. We present the design of a scalable, automated evaluation platform including of a VM cluster that can evaluate the generated code efficiently in various metrics that lead to various observations to facilitate selecting, evaluating, and optimizing code generation models. In the future, we plan to tackle longer, more complex YAML files with AI agents that break the goal into sub-tasks such as AutoGPT~\cite{autogpt}, extend \myname{} to more cloud applications and programming languages, as well as to experiment with state-of-the-art prompting methods such as chain-of-thought~\cite{wei2022chain}, retrieval-based LLM~\cite{lewis2020retrieval}, as well as fine-tuning techniques~\cite{gu2021ppt}.

\newpage
\bibliographystyle{mlsys2024}
\bibliography{references}

\clearpage


\section*{Supplementary material}

Table 8: Statistics of YAML files in top-100 most starred cloud native applications.

| Repo Name | Github Stars | Total Files | YAML Files | Repo Name | Github Stars | Total Files | YAML Files | Repo Name | Github Stars | Total Files | YAML Files |
|---|---|---|---|---|---|---|---|---|---|---|---|
| GitLab | 23368 | 58372 | 4721 | Dgraph | 19620 | 2231 | 161 | Terraform | 38875 | 5704 | 36 |
| Kubernetes | 101881 | 29662 | 4715 | Salt Project | 13513 | 7242 | 153 | Flink | 21993 | 27228 | 30 |
| Elastic | 65213 | 35747 | 3143 | Docker Compose | 30543 | 466 | 147 | Apollo | 28360 | 1512 | 28 |
| GraphQL | 30135 | 13667 | 2169 | Vitess | 16897 | 5579 | 142 | gVisor | 14172 | 3723 | 26 |
| Istio | 33694 | 6261 | 2081 | containerd | 14857 | 6523 | 138 | Sentinel | 21422 | 3487 | 25 |
| Ansible | 58659 | 7236 | 1914 | Serverless | 45187 | 1805 | 131 | go-zero | 25550 | 1382 | 22 |
| ShardingSphere | 18807 | 21945 | 1632 | CockroachDB | 27828 | 18499 | 118 | Seata | 24226 | 3904 | 21 |
| llvm | 21975 | 148442 | 1202 | k3s | 24517 | 750 | 97 | Packer | 14612 | 1450 | 20 |
| Argo | 14145 | 4172 | 1118 | Logstash | 13639 | 3835 | 88 | Wasmer | 16300 | 2007 | 19 |
| Skaffold | 14219 | 16345 | 1044 | Apache Spark | 36800 | 24415 | 85 | Portainer | 26644 | 3063 | 19 |
| Kubespray | 14472 | 2093 | 900 | Kong | 35947 | 1888 | 75 | Golang | 114620 | 14022 | 18 |
| SkyWalking | 22442 | 5999 | 802 | SST | 17715 | 4683 | 73 | SOPS | 13823 | 190 | 18 |
| Cilium | 16516 | 19972 | 780 | Rust | 85579 | 46998 | 69 | Redis | 61572 | 1679 | 16 |
| MongoDB | 24425 | 49784 | 743 | gRPC | 39066 | 12629 | 68 | kratos | 21387 | 861 | 16 |
| Backstage | 23285 | 12300 | 613 | Vault | 27546 | 9175 | 66 | NATS | 24451 | 580 | 16 |
| Grafana Loki | 20163 | 15520 | 554 | DragonflyDB | 21064 | 615 | 64 | Zig | 26009 | 16173 | 15 |
| Helm | 24953 | 1784 | 540 | Consul | 26921 | 13084 | 62 | Jenkins | 21453 | 13139 | 15 |
| Envoy | 22759 | 13470 | 520 | Keycloak | 17472 | 14535 | 59 | Apache Hadoop | 13858 | 9562 | 14 |
| Pulumi | 17622 | 8179 | 467 | Presto | 15087 | 13493 | 57 | Dubbo | 39400 | 5399 | 14 |
| Teleport | 14225 | 8884 | 419 | InfluxData | 26133 | 2007 | 56 | TiDB | 34880 | 6235 | 14 |
| Traefik | 44719 | 1870 | 339 | ORY Hydra | 14434 | 2556 | 56 | OpenFaaS | 23512 | 1100 | 14 |
| minikube | 27261 | 2368 | 316 | OpenAPI | 27136 | 181 | 55 | emscripten | 24266 | 9596 | 11 |
| SlimToolkit | 17269 | 6545 | 305 | Sentry | 35169 | 14388 | 54 | OpenCV | 71360 | 8613 | 10 |
| Prometheus | 49987 | 1389 | 255 | TDengine | 21762 | 4620 | 51 | Caddy | 49844 | 465 | 9 |
| Grafana | 57207 | 15782 | 242 | Jaeger | 18318 | 1469 | 48 | Apache bRPC | 15290 | 1632 | 9 |
| Podman | 19128 | 10589 | 203 | MinIO | 40904 | 1391 | 46 | Firecracker | 22578 | 822 | 8 |
| ClickHouse | 30874 | 27331 | 200 | Zipkin | 16425 | 1076 | 43 | Nacos | 27577 | 3501 | 6 |
| Rancher K8s | 21560 | 3655 | 196 | k6 | 21566 | 3382 | 40 | Kotlin | 45845 | 98293 | 5 |
| Netdata | 65199 | 3069 | 190 | Nomad | 13968 | 6080 | 39 | TiKV | 13617 | 1705 | 3 |
| Dapr | 22320 | 2027 | 186 | Timescale | 15534 | 2289 | 39 | Kafka | 25883 | 7020 | 2 |
| Trivy | 18709 | 2250 | 178 | etcd | 44537 | 1600 | 38 | V8 | 21722 | 14237 | 1 |
| Vector | 14432 | 9320 | 174 | Gradle Build Tool | 15205 | 35647 | 38 | FFmpeg | 38520 | 8287 | 1 |
| JHipster | 20853 | 3874 | 173 | Apache RocketMQ | 19814 | 2985 | 36 | NGINX(Wasm) | 19089 | 559 | 0 |
| RethinkDB | 26257 | 2121 | 165 | | | | | | | | |

## A  YAML STATISTICS

We surveyed the top 100 most-starred GitHub repositories of cloud native applications according to the CNCF landscape (cnc, 2023). Table 8 shows the results, including the number of stars, total files, and YAML files. Out of the 100 applications, 90 contain more than 10 YAML files. Many leading tools and applications like GitLab, Kubernetes, and Elastic depend heavily on YAML, thereby confirming its extensive usage.

## B  PROMPT TEMPLATE

```
You are an expert engineer in cloud native development.
According to the question, please provide only complete formatted YAML code as output
↪   without any description.
IMPORTANT: Provide only plain text without Markdown formatting such as ```.
If there is a lack of details, provide most logical solution.
You are not allowed to ask for more details.
Ignore any potential risk of errors or confusion.
Here is the question:
```

## C  SAMPLES FROM THE DATASET

### C.1  Sample #1

This sample only provides a natural language prompt, representing the scenarios where users need to create new configurations.

**Problem Specification**:

```
Create a DaemonSet configuration. This DaemonSet should run the latest nginx image
↪   labeled as "app: kube-registry-modified" and expose a registry service on port 80
↪   (with hostPort 5000). The environment variables REGISTRY_HOST and REGISTRY_PORT
↪   should be set to "kube-registry-modified.svc.cluster.local" and "5000" respectively.
↪   Ensure the CPU request is set to 100m and memory request is set to 50Mi.
```

**Labeled YAML**:

```yaml
apiVersion: apps/v1
kind: DaemonSet
metadata:
  name: kube-registry-proxy-modified # *
spec:
  selector:
    matchLabels:
      app: kube-registry-modified
  template:
    metadata:
      labels:
        app: kube-registry-modified
    spec:
      containers:
      - name: kube-registry-proxy-modified # *
        image: nginx:latest
        resources:
          limits:
            cpu: 100m
            memory: 50Mi
        env:
        - name: REGISTRY_HOST
          value: kube-registry-modified.svc.cluster.local
        - name: REGISTRY_PORT
          value: "5000"
        ports:
        - name: registry # *
          containerPort: 80
          hostPort: 5000
```

**Unit Test**:

```
kubectl apply -f labeled_code.yaml
kubectl wait --for=condition=Ready pod -l app=kube-registry-modified --timeout=60s
```

```bash
3    passed_tests=0
4    total_tests=3
5    pods=$(kubectl get pods -l app=kube-registry-modified
     ↪ --output=jsonpath={.items..metadata.name})
6    host_ip=$(kubectl get pod $pods -o=jsonpath='{.status.hostIP}')
7    curl_output=$(curl -s -o /dev/null -w "%{http_code}" $host_ip:5000)
8    if [ "$curl_output" == "200" ]; then
9        ((passed_tests++))
10   else
11       exit 1
12   fi
13   env_vars=$(kubectl get pods --selector=app=kube-registry-modified
     ↪ -o=jsonpath='{.items[0].spec.containers[0].env[*].name}')
14   if [[ $env_vars == *"REGISTRY_HOST"* && $env_vars == *"REGISTRY_PORT"* ]]; then
15       ((passed_tests++))
16   fi
17   cpu_limit=$(kubectl get pod $pods
     ↪ -o=jsonpath='{.spec.containers[0].resources.limits.cpu}')
18   memory_limit=$(kubectl get pod $pods
     ↪ -o=jsonpath='{.spec.containers[0].resources.limits.memory}')
19   if [ "$cpu_limit" == "100m" ] && [ "$memory_limit" == "50Mi" ]; then
20       ((passed_tests++))
21   fi
22   if [ $passed_tests -eq $total_tests ]; then
23       echo unit_test_passed
24   fi
```

### C.2 Sample #2

This sample provides a context YAML and seeks functionality extensions.

**Problem Specification**:

```
Given the following YAML, please help me create a service with load balancer that uses
↪ the nginx selector, exposed on port 80.
It should be accessible via browser.
```
```yaml
1    apiVersion: apps/v1
2    kind: Deployment
3    metadata:
4      name: nginx-deployment
5    spec:
6      replicas: 3
7      selector:
8        matchLabels:
9          app: nginx
10     template:
11       metadata:
12         labels:
13           app: nginx
14       spec:
15         containers:
16         - name: nginx-container
```

```
17            image: nginx:latest
18            ports:
19            - containerPort: 80
```

**Labeled YAML**:

```yaml
1  apiVersion: v1
2  kind: Service
3  metadata:
4    name: nginx-service    # *
5  spec:
6    selector:
7      app: nginx
8    ports:
9    - name: http
10     port: 80
11     targetPort: 80
12   type: LoadBalancer
```

**Unit Test**:

```
1  echo "apiVersion: apps/v1
2  kind: Deployment
3  [... the same as the YAML context in problem specification, omitted by brevity]
4         - containerPort: 80" | kubectl apply -f -
5  kubectl wait --for=condition=ready deployment --all --timeout=15s
6  kubectl apply -f labeled_code.yaml
7  sleep 15
8  kubectl get svc
9  timeout -s INT 8s minikube service nginx-service > bash_output.txt 2>&1
10 cat bash_output.txt
11 grep "Opening service default/nginx-service in default browser..." bash_output.txt &&
       echo unit_test_passed
```

### C.3 Sample #3

This is a debugging query sourced from StackOverflow. The raw YAML is included along with the error report. Typically, the response includes an error analysis unless restricted by the prompt template.

**Problem Specification**:

```
Given the following YAML which is not functionally correct:
1  apiVersion: networking.k8s.io/v1
2  kind: Ingress
3  metadata:
4    name: test-ingress
5  annotations:
6      nginx.ingress.kubernetes.io/rewrite-target: /
```

```yaml
7  spec:
8    rules:
9    - http:
10       paths:
11       - path: /
12         backend:
13           serviceName: test-app
14           servicePort: 5000
```

When executing it, it would report the error:

```
Error from server (BadRequest): error when creating "wrong.yaml": Ingress in version "v1"
↪   cannot be handled as a Ingress: strict decoding error: unknown field "annotations",
↪   unknown field "spec.rules[0].http.paths[0].backend.serviceName", unknown field
↪   "spec.rules[0].http.paths[0].backend.servicePort"

Please debug it to make it valid. Please provide the entire YAML.
```

**Labeled YAML**:

```yaml
1  apiVersion: networking.k8s.io/v1
2  kind: Ingress
3  metadata:
4    name: minimal-ingress
5    annotations:
6      nginx.ingress.kubernetes.io/rewrite-target: /
7  spec:
8    rules:
9    - http:
10       paths:
11       - path: /
12         pathType: Prefix
13         backend:
14           service:
15             name: test-app
16             port:
17               number: 5000
```

**Unit Test**:

```
1  kubectl apply -f labeled_code.yaml
2  kubectl wait --namespace default --for=condition=SYNCED ingress --all --timeout=15s
3  kubectl describe ingress minimal-ingress | grep "test-app:5000" && echo unit_test_passed
```

# D PROMPTS FOR DATA AUGMENTATION

## D.1 Prompt for question simplification

```
You are a helpful assistant trying to repharse INPUT paragraph that includes a question.
↪   Don't answer the question directly. Instead, please rewrite the question to make it
↪   as concise as possible, using abbrevations but include necessary informations. The
↪   question does not have to be a complete sentence. Don't modify code surrounded by ```.
↪   Let me first provide a few examples.

Example #1:
INPUT: Craft a yaml file to define a Kubernetes LimitRange. Containers within the cluster
↪   should have a default CPU request of 100m and a memory request of 200Mi. Any Pod
↪   created should not exceed a maximum CPU usage of 150m or a memory usage of 250Mi.
OUTPUT: Create a yaml for a k8s LimitRange. Containers CPU/Mem default to 100m/200Mi,
↪   pods max to 150m/250Mi.

Example #2:
INPUT: Please write a YAML file that defines firstly a Service and then a Deployment. The
↪   Deployment runs a single MySQL instance using the latest image on port 3306, with the
↪   environment MYSQL_ROOT_PASSWORD=password. The deployment should also define a volume
↪   mount for /var/lib/mysql, and then creates a PersistentVolumeClaim that matches the
↪   existing one. The Service simply expose the deployment on its port. All potential
↪   names should be mysql and labels should be app: mysql.
OUTPUT: Write a YAML defining a Service & Deployment. Deployment runs a MySQL instance on
↪   port 3306, env MYSQL_ROOT_PASSWORD=password, volume mount /var/lib/mysql, creates a
↪   matching PersistentVolumeClaim. Service exposes the deployment on its port. Using
↪   names mysql & labels app: mysql.

Example #3:
INPUT:
Is there a way to provide environment variables from a file in Kubernetes, for example
↪   when defining a pod, instead of hardcoding them like this:
```
```yaml
apiVersion: v1
kind: Pod
metadata:
  labels:
    context: docker-k8s-lab
    name: mysql-pod
  name: mysql-pod
spec:
  containers:
    - env:
        - name: MYSQL_USER
          value: mysql
        - name: MYSQL_PASSWORD
          value: mysql
        - name: MYSQL_DATABASE
          value: sample
        - name: MYSQL_ROOT_PASSWORD
          value: supersecret
      image: "mysql:latest"
      name: mysql
```

```yaml
21       ports:
22         - containerPort: 3306
```

Please provide the entire YAML, supposing there is a Secret named mysql-secret, with all
↪   values in it.
OUTPUT:
Can k8s use env var from a file instead of hardcoding? Given:

```yaml
1  apiVersion: v1
2  kind: Pod
3  metadata:
4    labels:
5      context: docker-k8s-lab
6      name: mysql-pod
7    name: mysql-pod
8  spec:
9    containers:
10     - env:
11        - name: MYSQL_USER
12          value: mysql
13        - name: MYSQL_PASSWORD
14          value: mysql
15        - name: MYSQL_DATABASE
16          value: sample
17        - name: MYSQL_ROOT_PASSWORD
18          value: supersecret
19       image: "mysql:latest"
20       name: mysql
21       ports:
22         - containerPort: 3306
```

Assume a Secret named mysql-secret with all values. Provide the full YAML.

Example #4:
INPUT:
I'm working with the bookinfo application in our Istio setup.
I recall there was a DestinationRule specifically for the ratings service in the prod
↪   namespace, which ensures traffic is load balanced using the LEAST_REQUEST strategy.
Please provide me the exact configuration for that.
OUTPUT:
Provide exact config for DestinationRule for ratings service in prod namespace using
↪   LEAST_REQUEST lb strategy in Istio/bookinfo app setup.

Example #5:
INPUT:
I need a Istio destination rule YAML set up for the bookinfo application's ratings
↪   service in the prod namespace.
This rule had the main traffic load balanced using the LEAST_REQUEST strategy.
Additionally, there was a specific subset named testversion using version v3 labels, and
↪   for this subset, the traffic was load balanced with a ROUND_ROBIN approach.
Please provide me the entire YAML configuration for this.
OUTPUT:
Provide Istio DestinationRule YAML for bookinfo app's ratings service in prod ns. Main
↪   traffic uses LEAST_REQUEST lb, subset "testversion" uses labels v3 and ROUND_ROBIN lb
↪   strategy.

```
That's all the examples I have. Now let's working on the following input:
INPUT: {question}
```

### D.2 Prompt for question translation

```
You are a helpful assistant trying to translate INPUT paragraph that includes a question
↪   into fluent Chinese in a developer's tone.
Don't answer the question directly. Instead, please just translate the question for a
↪   Chinese developer to answer. The question does not have to be a formal and complete
↪   sentence. Don't modify code surrounded by ```. Let me first provide a few examples.

Example #1:
INPUT: Craft a yaml file to define a Kubernetes LimitRange. Containers within the cluster
↪   should have a default CPU request of 100m and a memory request of 200Mi. Any Pod
↪   created should not exceed a maximum CPU usage of 150m or a memory usage of 250Mi.

OUTPUT: 写一个 yaml 来定义 Kubernetes LimitRange。 集群里的容器的默认 CPU 请求是 100m ，默认内存
↪   请求是 200Mi 。 创建的任何 Pod 的 CPU 使用量最大不应超过 150m ，内存使用最大不应超过 250Mi。

Example #2:
INPUT: Please write a YAML file that defines firstly a Service and then a Deployment. The
↪   Deployment runs a single MySQL instance using the latest image on port 3306, with the
↪   environment MYSQL_ROOT_PASSWORD=password. The deployment should also define a volume
↪   mount for /var/lib/mysql, and then creates a PersistentVolumeClaim that matches the
↪   existing one. The Service simply expose the deployment on its port. All potential
↪   names should be mysql and labels should be app: mysql.
OUTPUT: 请写一个 YAML ，先定义 Service，再定义 Deployment 。 该 Deployment 使用端口 3306 上的最
↪   新映像运行单个 MySQL 实例，环境为 MYSQL_ROOT_PASSWORD=password。 Deployment还应该为
↪   /var/lib/mysql 定义一个卷挂载，然后创建一个与现有卷匹配的 PersistentVolumeClaim。 该 Service
↪   只是在其端口上公开部署。 所有潜在名称定为 mysql,标签定为 app:mysql。

Example #3:
INPUT:
Is there a way to provide environment variables from a file in Kubernetes, for example
↪   when defining a pod, instead of hardcoding them like this:
```
```yaml
apiVersion: v1
kind: Pod
metadata:
  labels:
    context: docker-k8s-lab
    name: mysql-pod
  name: mysql-pod
spec:
  containers:
    - env:
        - name: MYSQL_USER
          value: mysql
        - name: MYSQL_PASSWORD
          value: mysql
        - name: MYSQL_DATABASE
```

```yaml
16            value: sample
17          - name: MYSQL_ROOT_PASSWORD
18            value: supersecret
19        image: "mysql:latest"
20        name: mysql
21        ports:
22          - containerPort: 3306
```
Please provide the entire YAML, supposing there is a Secret named mysql-secret, with all
↪  values in it.
OUTPUT:
有没有办法从 Kubernetes 中的文件提供环境变量，比如在定义 pod 时，而不是像下面这样对它们进行硬编码：

```yaml
1  apiVersion: v1
2  kind: Pod
3  metadata:
4    labels:
5      context: docker-k8s-lab
6      name: mysql-pod
7    name: mysql-pod
8  spec:
9    containers:
10     - env:
11         - name: MYSQL_USER
12           value: mysql
13         - name: MYSQL_PASSWORD
14           value: mysql
15         - name: MYSQL_DATABASE
16           value: sample
17         - name: MYSQL_ROOT_PASSWORD
18           value: supersecret
19       image: "mysql:latest"
20       name: mysql
21       ports:
22         - containerPort: 3306
```
请提供整个 YAML，假设有一个名为 mysql-secret 的 Secret，其中包含所有值。

Example #4:
INPUT:
I'm working with the bookinfo application in our Istio setup.
I recall there was a DestinationRule specifically for the ratings service in the prod
↪  namespace, which ensures traffic is load balanced using the LEAST_REQUEST strategy.
Please provide me the exact configuration for that.
OUTPUT:
我正在 Istio 配置中使用 bookinfo 应用。
我记得有一个专门用于生产命名空间中的 ratings 服务的 DestinationRule，它确保使用 LEAST_REQUEST 策略
↪   进行流量负载平衡。
请为此提供确切的配置。

Example #5:
INPUT:
I need a Istio destination rule YAML set up for the bookinfo application's ratings
↪   service in the prod namespace.
This rule had the main traffic load balanced using the LEAST_REQUEST strategy.

```
Additionally, there was a specific subset named testversion using version v3 labels, and
↪  for this subset, the traffic was load balanced with a ROUND_ROBIN approach.
Please provide me the entire YAML configuration for this.
OUTPUT:
我需要为 prod namespace 中的 bookinfo 应用的评级服务配置一个 Istio destination rule YAML。
该规则使用 LEAST_REQUEST 策略对主要流量进行负载平衡。
此外，还有一个名为 testversion 的特定子集，使用版本 v3 标签，对于该子集，流量通过 ROUND_ROBIN 方法进
↪  行负载平衡。
请为此提供完整的 YAML 配置。

That's all the examples I have. Now let's working on the following input:
INPUT: {question}
```

# E  PERFORMANCE ANALYSIS ON DIFFERENT AFFECTING FACTORS

Table 9: Performance analysis on different affecting factors

|  | Category | | | Code context | | Reference code length | | | Question token number | | |
|---|---|---|---|---|---|---|---|---|---|---|---|
|  | Kubernetes | Envoy | Istio | w/ code | w/o code | [0, 15) | [15, 30) | ≥30 | [0, 50) | [50, 100) | ≥100 |
| gpt-4 | 0.601 | 0.1 | 0.385 | 0.51 | 0.547 | 0.625 | 0.616 | 0.237 | 0.619 | 0.563 | 0.427 |
| gpt-3.5 | 0.466 | 0.122 | 0.385 | 0.429 | 0.416 | 0.534 | 0.477 | 0.169 | 0.476 | 0.423 | 0.378 |
| PaLM-2-bison | 0.406 | 0.05 | 0.231 | 0.354 | 0.358 | 0.455 | 0.413 | 0.118 | 0.44 | 0.345 | 0.309 |
| Llama-2-70b-chat | 0.099 | 0.049 | 0 | 0.082 | 0.095 | 0.216 | 0.058 | 0.013 | 0.167 | 0.077 | 0.045 |
| Llama-2-13b-chat | 0.085 | 0.049 | 0 | 0.061 | 0.089 | 0.125 | 0.081 | 0.013 | 0.143 | 0.063 | 0.045 |
| Wizardcoder-34b-v1.0 | 0.067 | 0.05 | 0.231 | 0.068 | 0.074 | 0.159 | 0.052 | 0.013 | 0.131 | 0.049 | 0.055 |
| Llama-2-7b-chat | 0.039 | 0.05 | 0 | 0.054 | 0.026 | 0.08 | 0.029 | 0.013 | 0.036 | 0.042 | 0.036 |
| Wizardcoder-15b-v1.0 | 0.032 | 0.049 | 0.077 | 0.061 | 0.016 | 0.045 | 0.041 | 0.013 | 0.012 | 0.035 | 0.054 |
| Llama-7b | 0.035 | 0.05 | 0 | 0.054 | 0.021 | 0.057 | 0.035 | 0.013 | 0.024 | 0.042 | 0.036 |
| Llama-13b-lora | 0.021 | 0.049 | 0 | 0.041 | 0.011 | 0.034 | 0.017 | 0.026 | 0.012 | 0.035 | 0.054 |
| Codellama-13b-instruct | 0.011 | 0.05 | 0 | 0.007 | 0.021 | 0.034 | 0.006 | 0.013 | 0 | 0.021 | 0.018 |
| Codellama-7b-instruct | 0.007 | 0.049 | 0.077 | 0.014 | 0.016 | 0.034 | 0.006 | 0.013 | 0.012 | 0.007 | 0.027 |


\end{document}